\documentclass[
aps,
prd,
reprint,
tightenlines,
superscriptaddress,
nofootinbib,
]{revtex4-2}

\usepackage{booktabs}
\usepackage{makecell}
\usepackage{braket}
\usepackage{amsmath}
\usepackage{eufrak}
\usepackage{graphicx}
\usepackage{subfigure}
\usepackage{grffile}
\usepackage{float}
\usepackage[a4paper, total={6.5in, 9in}]{geometry}
\usepackage[colorlinks,
linkcolor=blue,
anchorcolor=blue,
citecolor=blue
]{hyperref}
\usepackage{diagbox}

\usepackage{xcolor}
\usepackage{comment}
\usepackage{tgtermes}  

\newcommand{\rds}{\rho_{\rm DS}}

\newcommand{\bea}{\begin{eqnarray}}
\newcommand{\eea}{\end{eqnarray}}

\newcommand{\bk}{{\bf k}}
\newcommand{\bx}{{\bf x}}
\newcommand{\D}{{\rm d}}
\newcommand{\mpl}{M_\mathrm{Pl}}
\newcommand{\DS}{\mathrm{DS}}

\xdef\figsizeTwo{7.8cm}
\xdef\figsizeOne{7.5cm}

\date{\small\itshape Last update: \today}
\begin{document}
\title{A New Origin of the Big Bang from Dark-Sector-Induced Vacuum Decay and Its Gravitational-Wave Signal}
\author{Haipeng An}
\email{anhp@mail.tsinghua.edu.cn}
\affiliation{Department of Physics, Tsinghua University, Beĳing 100084, China}
\affiliation{Center for High Energy Physics, Tsinghua University, Beijing 100084, China}

\author{Tingyu Li}
\email{lity22@mails.tsinghua.edu.cn}
\affiliation{Department of Physics, Tsinghua University, Beĳing 100084, China}

\begin{abstract}

We propose a novel scenario for the onset of the thermal Big Bang. In this framework, the inflaton transfers its energy exclusively into a dark sector, leaving the Standard Model (SM) sector temporarily trapped in a false vacuum. As the Hubble expansion rate rapidly decreases, the SM phase transition eventually completes, and the standard thermal Big Bang era commences upon the thermalization of the highly energetic bubble walls. We demonstrate that the large Lorentz boost of these bubble walls, combined with their Hubble-scale macroscopic size, generates distinctive gravitational-wave signatures from the SM vacuum decay. This stochastic gravitational-wave background provides a powerful new probe of the early Universe's expansion history, with a present-day energy density fraction that can reach $\Omega_{\text{GW}} \sim 3\times10^{-8}$.

\end{abstract}
\maketitle	

%
{\noindent\itshape Introduction}---
The cosmological constant problem is one of the deepest puzzles in fundamental physics. At present, the only broadly accepted framework for addressing it invokes the anthropic principle~\cite{Weinberg:1987dv,Weinberg:1988cp}. For this idea to be viable, however, a landscape of vacua must exist.
Another fundamental question concerns the nature of dark matter. The absence of convincing signals in direct and indirect searches for particle dark matter may indicate that dark matter interacts with Standard Model (SM) particles only through gravity. This possibility naturally suggests the existence of one or more dark sectors (DSs), in which dark matter resides and which are coupled to the SM sector purely gravitationally.

Motivated by these considerations, we propose a new scenario for the origin of the hot Big Bang. We assume that, immediately after inflation, the inflaton transfers all of its energy into one or more DSs, while the SM sector remains trapped in a metastable vacuum ${\bf v}_1$. Subsequently, quantum tunneling can trigger the nucleation of vacuum bubbles, allowing the SM sector to transition to a nearby vacuum ${\bf v}_2$ with lower vacuum energy. The bubble nucleation rate per unit physical volume is given by~\cite{Coleman:1977py,Callan:1977pt}
\bea\label{eq:GammaV}
\frac{\Gamma}{\mathcal V}=m^4 e^{-S},
\eea
where $m$ is the characteristic energy scale of the phase transition, and $S$ is the bounce action associated with the potential barrier.

\begin{figure}
    \centering
    \hspace*{-0.5cm}    
    \includegraphics[width=\figsizeOne]{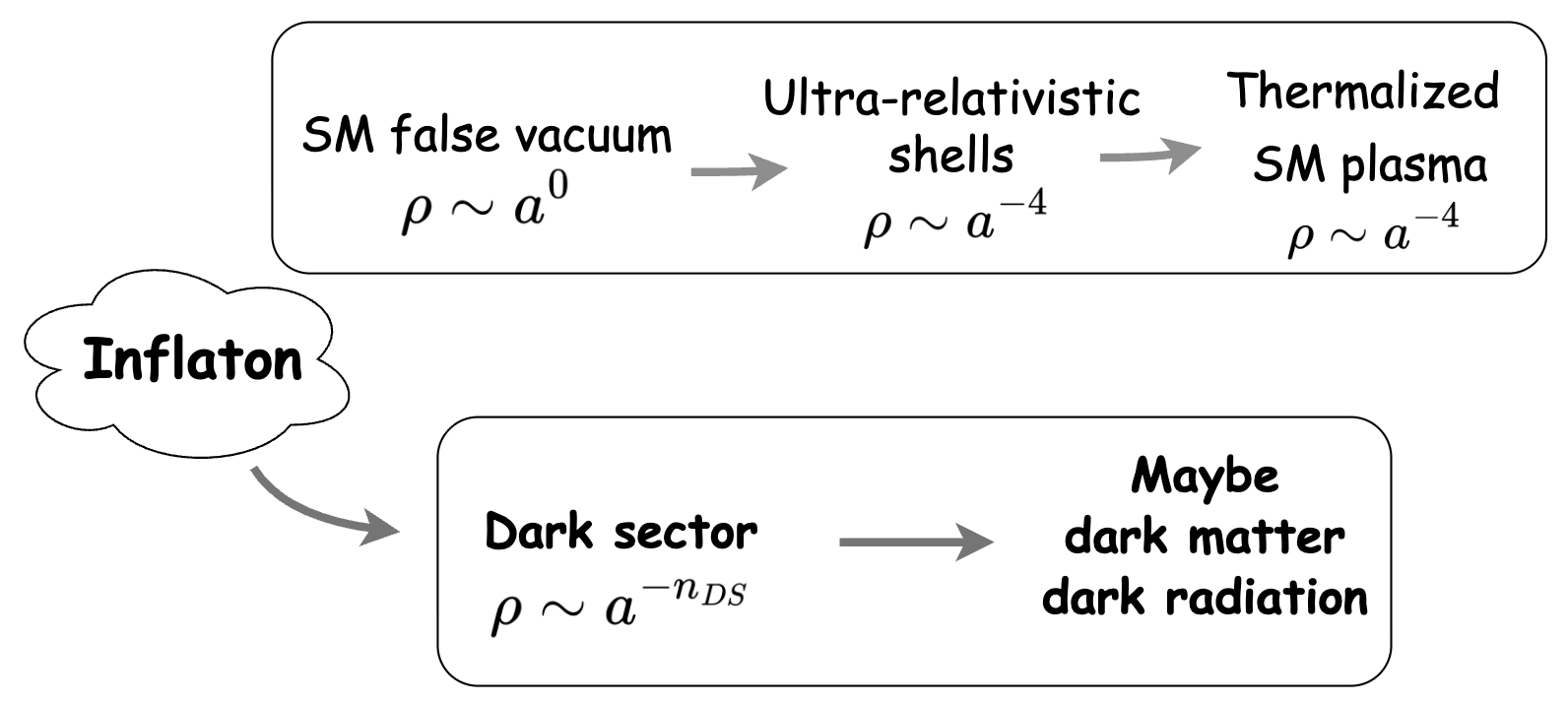}
    \caption{
    Illustration of DS-triggered SM vacuum decay.
    }
    \label{fig:strategy}
\end{figure}
\bigskip

\noindent{\it Evolution of the energy density of the Universe}---We further assume that the DS has an equation of state such that its energy density scales as \(\rho_{\rm DS}\propto a^{-n_{\rm DS}}\), and that it dominates the expansion of the Universe before the SM-sector phase transition completes. This assumption avoids the overproduction of primordial black holes (PBHs). The details of our estimation of the PBH constraints are presented in (SUPP). Current observations require any dark-radiation component to contribute no more than about \(10\%\) of the CMB energy density~\cite{DESI:2024mwx,Allali:2024cji}, implying \(n_{\rm DS}>4\). One realization of this scenario occurs if, after inflation, the inflaton transfers its potential energy into the kinetic energy of itself or of other fields, leading to a period of kination domination (KD)~\cite{Spokoiny:1993kt,Joyce:1996cp,Peebles:1998qn,Ellis:2020krl}, for which \(n_{\rm DS}=n_K=6\). The Universe may also enter KD after an intermediate matter-dominated stage~\cite{Co:2021lkc}. More generally, the inflaton may deposit its energy into multiple dark sectors, each with a different equation of state. Therefore, in the following discussion, we consider generic values of \(n_{\rm DS}\).

The Hubble expansion rate $H$ decreases monotonically after inflation, since $H^2$ is proportional to the total energy density, $\rho_{\rm tot}=\rho_{\rm SM}+\rds$~\cite{Friedmann:1922,Friedmann:1924}. Suppose that initially $H^4>\Gamma/{\cal V}$, so that the SM sector remains trapped in the metastable vacuum ${\bf v}_1$. As the Universe expands, however, $H$ eventually reaches a critical value $H_c$ defined by
\bea\label{eq:Hc}
H_c^4 = \Gamma/{\cal V} \ .
\eea
At this point, the SM-sector phase transition begins to complete. After bubble collisions, the vacuum energy is converted into the kinetic energy of expanding shell-like configurations, whose energy density subsequently redshifts as $a^{-4}$, like radiation~\cite{An:2026hiq}. Thus, the evolution of the total energy density of the Universe can be written as
\bea
\rho_{\rm tot} = \left\{ \begin{array}{cc} 
L + \rho_c (a/a_c)^{-n_{\rm DS}}\ , ~&~ a \leq a_c \,, \\ 
L(a/a_c)^{-4} + \rho_c (a/a_c)^{-n_{\rm DS}}\ , ~&~ a \geq a_c \, .
\end{array}
\right.
\eea
After thermalization, the SM sector enters the standard hot Big Bang phase. Figure~\ref{fig:strategy} illustrates this scenario for the origin of the Big Bang. In what follows, we discuss the properties and detectability of the gravitational waves (GWs) produced during this phase transition.

\bigskip

{\noindent\itshape Bubble collisions}---
%
%
When the occupation fraction of the new vacuum \({\bf v}_2\) becomes of order unity, Eq.~\eqref{eq:Hc} is satisfied. The typical bubble radius at the completion of the phase transition, \(R_c\), is then of order \(H_c^{-1}\).
Before collisions, the bubble walls are accelerated by the vacuum-energy difference. The boost factor of the walls immediately before the first collisions can be estimated as
\bea\label{eq:gamma}
\gamma \sim \frac{L R_c^3}{\sigma R_c^2}
\sim \frac{m}{H_c}
\sim \frac{m\mpl}{(L+\rho_c)^{1/2}}
\sim \alpha^{1/2}\frac{\mpl}{m} \ ,
\eea
where we have used the generic relations \(L\sim m^4\) and \(\sigma\sim m^3\), with \(\sigma\) denoting the bubble-wall tension in its rest frame, and \(\alpha \equiv L/(L+\rho_c)\). Thus, throughout most of the parameter space of interest, \(\gamma\gg1\). Upon collision, the characteristic wall thickness in the bubble rest frame is \((\gamma m)^{-1}\sim \alpha^{-1/2}\mpl^{-1}\), i.e., close to the Planck scale.

We describe the bubbles by a scalar field \(\phi\) satisfying
\bea\label{eq:eom}
\ddot\phi + 3H\dot\phi - \nabla^2\phi + V'(\phi)=0 \ ,
\eea
where \(V'(\phi)\) denotes the derivative of the potential \(V\) with respect to \(\phi\). During bubble collisions, the different terms scale parametrically as
\[
\ddot\phi\sim\gamma^2,~~
\nabla^2\phi\sim\gamma^2,~~
3H\dot\phi\sim\gamma,~~
V'(\phi)\sim\gamma^0 .
\]
Therefore, for highly boosted walls, the kinetic and gradient terms dominate over the Hubble-friction and potential terms. As a result, the bubble walls largely pass through one another and form highly energetic shell-like configurations. The collision dynamics and energy-shell formation in the large-boost limit have been thoroughly discussed in Ref.~\cite{An:2026hiq}.

\bigskip

\noindent{\it Evolution of the high energy shells}--- 
Driven by non-linear scalar potentials and gauge interactions, the expanding collision shells inevitably evolve and thermalize. While a full numerical simulation of specific models is beyond the scope of this work, the underlying physical mechanisms guarantee efficient thermalization, as demonstrated by the following phenomenological arguments.

Consider the SM Higgs field as a benchmark. Following the initial bubble collisions, the highly boosted shells ($\gamma \sim 10^{15}$, as derived in Eq.~\eqref{eq:gamma}) consist predominantly of Higgs bosons. Crucially, by virtue of the Goldstone equivalence theorem~\cite{Cornwall:1974km}, the scattering channels $hh\rightarrow W_L W_L$ and $hh\rightarrow Z_L Z_L$ via Coulomb-like gauge exchange suffer no $\gamma$-suppression. In the large-$\gamma$ limit, the combined cross section for longitudinal vector boson production approaches a robust constant:
\begin{equation}
    \sigma_{hh\rightarrow WW,ZZ} = \frac{g_2^4(2+\cos^{-2}\theta_W)}{64\pi m_W^2} \ ,
\end{equation}
where $g_2$ is the $\text{SU}(2)_L$ weak coupling constant and $\theta_W$ is the Weinberg angle. 

Inside a high-energy shell dominated by the Higgs field, the rest-frame particle number density and width naturally scale as $m_h^3$ and $m_h^{-1}$, respectively. Consequently, each shell crossing converts a predictable fraction of Higgs bosons into gauge bosons, roughly $g_2^4/(4\pi) \sim 10^{-3}$. Once produced, these vector bosons rapidly decay into quarks and subsequently hadronize. Because the observer-frame decay rate, $\gamma^{-1} g_2^2 m_W / (4\pi)$, translates to a per-Hubble-time decay fraction of $H_c H^{-1} g_2^2 / (4\pi)$, the decay of the vector bosons is vastly more efficient than their initial production. The thermalization bottleneck is therefore the initial scalar conversion, not the subsequent decay.

As the system evolves, the dynamics are rapidly dominated by the newly generated hadrons. Hadronic cross sections are dictated by the strong interaction scale, which sits at $\Lambda_{\text{QCD}}^{-2} \sim \mathcal{O}(1)~\mu\text{b}$. These energetic collisions trigger massive parton showers, with multiplicities scaling efficiently as $\sqrt{E/\Lambda_{\text{QCD}}}$~\cite{Khoze:1996dn}. Given an enormous optical depth of $\sigma v n d \sim m_W^2/\Lambda_{\text{QCD}}^2 \gg 1$, the shells immediately become completely opaque to hadrons, ensuring that the strongly interacting sector thermalizes within just a handful of collisions. Because the overall thermalization time is bottlenecked by the roughly $1000$ shell crossings needed for the initial Higgs conversion, we can reliably estimate the reheating temperature by accounting for the cosmic redshift accumulated during this period. This yields a definitive reheating scale of approximately $10^{-3} m_W \sim 100$~MeV.

This physical picture readily extends to beyond-the-Standard-Model scenarios. For instance, consider a vacuum decay occurring within a Left-Right Symmetric Model~\cite{Mohapatra:1974gc,Senjanovic:1975rk}, where the bubble walls are composed of scalar fields that transition into right-handed gauge bosons ($W_R$). These $W_R$ bosons subsequently decay into light quarks with a rate of approximately $g_2^2 m_{W_R}/(4\pi)$. Assuming the scalar mass is comparable to $m_{W_R}$, the overall thermalization timescale is dictated by this $W_R$ decay rate. Consequently, it takes roughly $4\pi/g_2^2 \sim \mathcal{O}(100)$ collisions for the shells to fully thermalize, yielding a reheating temperature governed by $\Gamma_{W_R} \sim 10^{-2} m_{W_R}$. These distinct scenarios reveal a universal feature of phase transition dynamics: the reheating temperature is fundamentally dictated by the number of shell collisions required to transfer energy into the fastest-thermalizing—typically strongly interacting—sector. Whether governed by standard QCD or new strongly coupled dynamics in composite Higgs models, these robust estimates demonstrate that highly energetic expanding shells reliably thermalize after $\mathcal{O}(10^2)$ to $\mathcal{O}(10^3)$ collisions.

\bigskip

\noindent{\it GW production} --- 
The GWs are described by \(h_{ij}^{TT}\), the transverse-traceless part of the metric perturbation,
\[
d\tau^2 = a^2(\eta)\left[d\eta^2 - \left(\delta_{ij}+h_{ij}\right)dx^i dx^j\right],
\]
where \(\eta\) is the conformal time, defined by \(\D\eta = a^{-1}\D t\). The Fourier transform of \(h_{ij}^{TT}\) can be obtained using the Green's function method,
\bea\label{eq:hTT}
\tilde h_{ij}^{TT}(\eta,\bk)
&=&
\frac{16\pi G_N}{a(\eta)}
\int \D\eta' \nonumber\\
&&G(\eta,\eta',k)
a(\eta')
\tilde T_{ij}^{TT}(\eta',\bk) ,
\eea
where \(G(\eta,\eta',k)\) is the retarded Green's function for GWs in Fourier space, and \(\tilde T_{ij}^{TT}\) is the Fourier transform of the transverse-traceless part of the energy-momentum tensor. Since \(\rho_{\rm SM}\) and \(\rho_{\rm DS}\) evolve differently during the phase transition, \(G(\eta,\eta',k)\) has no closed-form expression and must be computed numerically. Nevertheless, its generic behavior is that it oscillates as a function of both \(k\eta\) and \(k\eta'\) once \(k\eta\) and \(k\eta'\) become larger than unity.


From the field configuration, we compute the Fourier modes of the transverse-traceless part of the scalar-field energy-momentum tensor. These can be written as a convolution of the Fourier transform of \(\phi\),
\bea
\tilde T^{TT}_{ij}(\mathbf{k})
=
\int\frac{d^3q}{(2\pi)^3}
\Lambda_{ijkl}\,
q_k k_l\,
\tilde \phi(\mathbf{q})\,
\tilde\phi^*(\mathbf{q}-\mathbf{k}) \ ,
\eea
where \(\Lambda_{ijkl}\) projects \(T_{ij}\) onto its transverse-traceless components. The \(q\)-integral is UV dominated, with the dominant contribution coming from the region \(q\sim \gamma m\). 
%
In this work, we take advantage of the large boost nature of the bubble walls to use the bulk-flow model~\cite{Konstandin:2017sat,Ellis:2020nnr} to simulate the evolution of $T_{ij}^{TT}$. In the following, we analyze qualitatively the strength, peak frequency, and shape of the GW signal, and leave the details of the simulations in the appendices.


From \(\tilde h_{ij}^{TT}\), we can directly construct the GW spectrum immediately after production,
\bea
\Omega_{\rm GW}^*
\equiv
\frac{\D\rho_{\rm GW}}{\rho_R \D\ln k}
\propto
k^3
\int \frac{\D\hat\bk}{4\pi}
\sum_{ij}
\left|
\partial_t \tilde h_{ij}^{TT}(t,\bk)
\right|^2 .
\eea
The present-day GW spectrum is then obtained as
\bea
\Omega^0_{\rm GW}
=
5.0\times10^{-5}\times
\left(\frac{106.75}{g_{\rm SM}}\right)^{1/3}
\Omega^*_{\rm GW} \ .
\eea
We present the details of the simulations in the End Matter (EM) and the Supplemental Material (SUPP). In the following, we analyze the qualitative features of \(\Omega^0_{\rm GW}\).

\bigskip

\noindent{\it GW strength} ---
In such a relativistic system, the GW energy density can be estimated from the gravitational potential energy density~\cite{Witten:1984rs,Hogan:1986dsh},
\bea
\rho_{\rm GW} \sim \frac{G_N M_B^2}{R^4}
\sim G_N L^2 H_c^{-2} \ ,
\eea
where \(G_N\) is Newton's constant. Here the bubble mass has been estimated as \(M_B \sim L R_c^3\), with \(L\) denoting the latent energy density released during the phase transition, and \(R_c\sim H_c^{-1}\) the characteristic bubble radius at collision.

In vacuum decay, immediately before collision, the released latent energy is stored in the highly boosted bubble walls. After collision, the walls pass through one another and form highly boosted energy shells with soliton-like field configurations. The corresponding SM energy density therefore behaves as radiation and redshifts as \(a^{-4}\). Consequently, once these highly boosted energy shells thermalize, the SM sector smoothly enters the standard hot Big Bang expansion phase.

Under this crude estimate, the present-day GW energy density can be written as
\bea
\Omega^0_{\rm GW}
\sim
\Omega_R \times \frac{\rho_{\rm GW}}{L}
\sim
\Omega_R \times \frac{L}{\rho_{\rm tot}}
=
\alpha \Omega_R \ ,
\eea
where we have used the Friedmann equation \(H^2 = 8\pi G_N \rho_{\rm tot}/3\). Here \(\rho_{\rm tot}\) is the total energy density of the Universe at the beginning of the phase transition, and, as usual, \(\alpha \equiv L/\rho_{\rm tot}\).

The peak amplitude of the GW spectrum can therefore be parameterized as
\bea
\left.
\frac{\D\Omega_{\rm GW}}{\D \ln k}
\right|_{\rm peak}
=
C_p\,\alpha\,\Omega_R \ ,
\eea
where, due to the expansion of the Universe during the phase transition, the coefficient \(C_p\) depends mildly on \(n\) and \(\alpha\). For several representative values of \(n_{\rm DS}\) and \(\alpha\), the corresponding values of \(C_p\) are shown in Table~\ref{tab:Cs}.
\begin{table}[htbp]
  \centering
  \caption{$C_p / C_k$ for different $n$ and $\alpha$}
  \label{tab:Cs}
  \begin{tabular}{c@{\hspace{0.5em}}c@{\hspace{0.5em}}c@{\hspace{0.5em}}c@{\hspace{0.5em}}c@{\hspace{0.5em}}c}
    \hline\hline             
    \diagbox[width=2.0em, height=2.0em]{$n$}{$\alpha$} &  0.02&  0.03&  0.04&  0.05&  0.06 \\
    \hline                    
    4 &0.011/4.6 & 0.012/4.7 & 0.011/5.2 & 0.011/5.6 & 0.012/5.9 \\
    5 & 0.010/4.0 & 0.010/4.5 & 0.011/4.8 & 0.011/5.1 & 0.011/5.3 \\
    6 & 0.010/3.8 & 0.011/4.1 & 0.010/4.5 & 0.010/4.6 & 0.010/4.9 \\
    \hline\hline              
  \end{tabular}
\end{table}

\bigskip

\noindent{\it GW peak frequency} ---
The peak GW frequency is determined by the characteristic bubble size, which is of order $H_c^{-1}$. Consequently, at the time of production, the peak physical wavenumber is given by $k^{\rm peak}_{\rm phys} = C_k H_c$, where $C_k$ is an ${\cal O}(1)$ parameter. The precise value of $C_k$ is model-dependent; we present its values for several benchmark scenarios in Table~\ref{tab:Cs}.

\bigskip

\noindent{\it GW spectroscopy} --- 
Since the Green's function $G(\eta,\eta',k)$ oscillates for $k\eta' \gg 1$, the $\eta'$ integral in Eq.~\eqref{eq:hTT} is effectively cut off at $\eta' \sim k^{-1}$. In the regime $k\eta \ll 1$, $G$ depends on the detailed composition of the energy density. Assuming $\rho_{\rm tot}\propto a^{-n}$, we have $G\sim k^{-1/2}\eta'^{1/2}(k\eta')^{-\nu}$, with $\nu = |1/2 - 2/(n-2)|$. Specifically, if $\rho_{\rm DS}$ decays sufficiently fast (such as in a KD regime), $\rho_{\rm SM}$ dominates the Universe soon after the phase transition completes. In this scenario, $n \simeq 4$, and then the IR limit of $G$ becomes roughly independent of $\eta'$. 
\begin{figure}
    \centering
    \hspace*{-0.5cm}
    \includegraphics[width=\figsizeTwo]{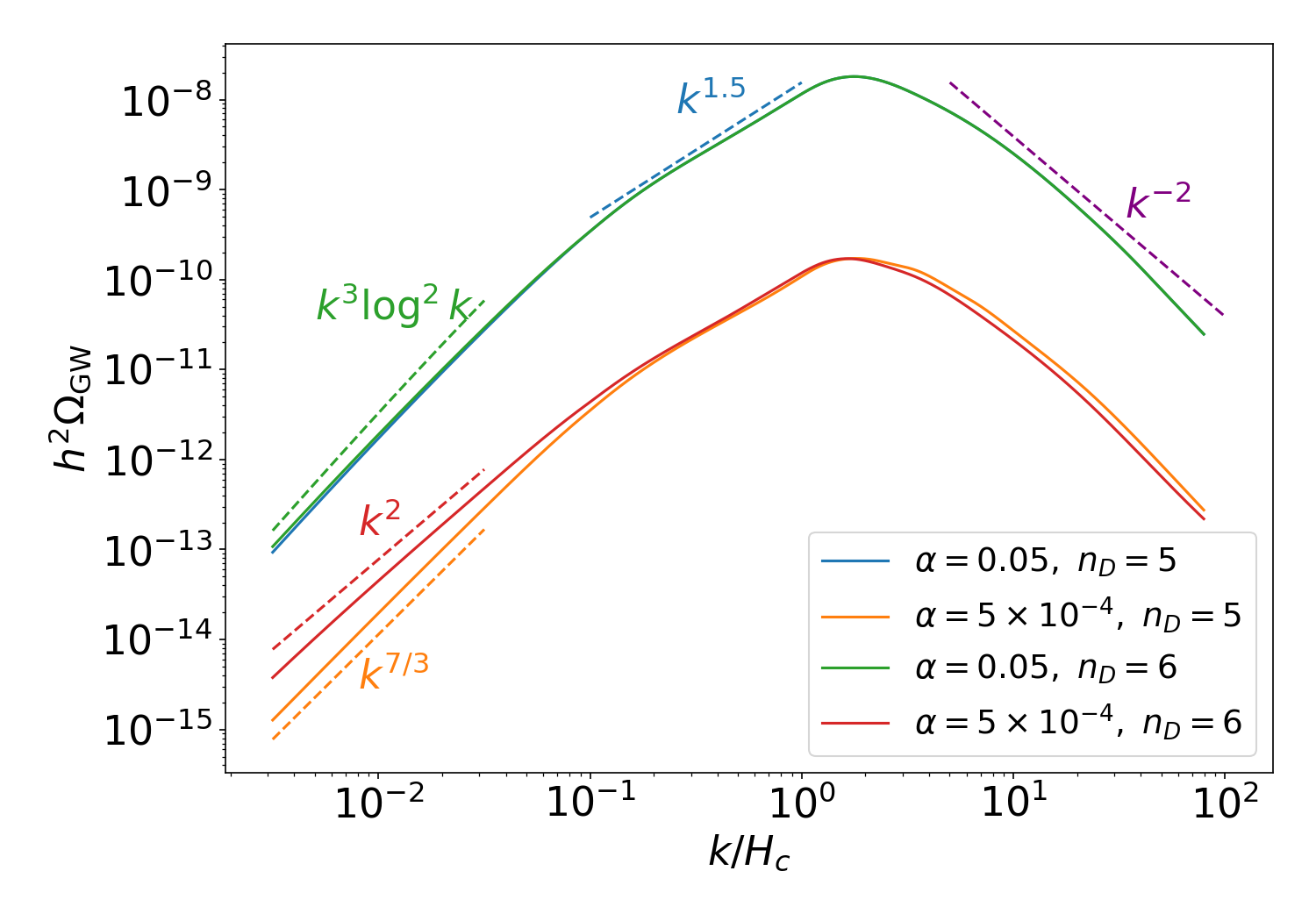}
    \caption{
    The shape of gravitational-wave energy density spectrum for different $n_{D}$ and $\alpha$. The IR scaling depends on the background evolution details through Eq.\eqref{eq:omega IR scaling}.}
    \label{fig:gw0}
\end{figure}

\begin{figure}[htb]
    \centering
    \hspace*{-0.5cm}
    \includegraphics[width=\figsizeTwo]{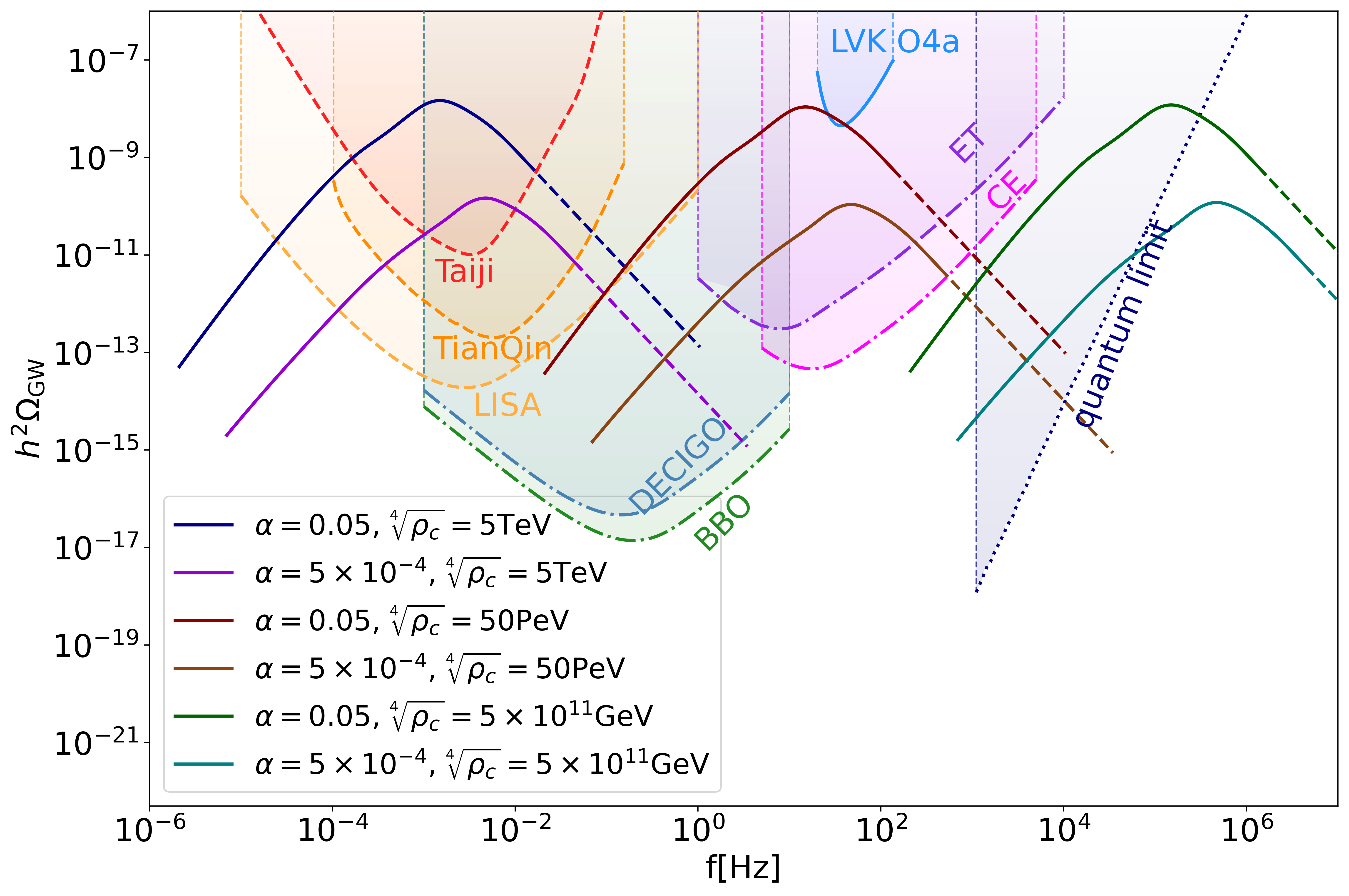}
    \caption{
    The gravitational-wave energy density spectrum from a dark-sector-induced vacuum phase transition for $n_{DS}=6$. Here, $\alpha$ is chosen to be $5\times10^{-2}$ and $5\times10^{-4}$, respectively. We consider three energy scales for the phase transition, with $\sqrt[4]{\rho_{i}}=5\mathrm{TeV}, 50\mathrm{PeV}, 5\times 10^{11}\mathrm{GeV}$. For comparison, we also show the sensitivity curve of LIGO--Virgo--KAGRA~\cite{LIGOScientific:2025kry}, along with those of several future gravitational-wave detectors. The sensitivity curves for ET, CE, LISA, DECIGO, and BBO are taken from Ref.\cite{Schmitz:2020syl}. Those for Taiji and TianQin are taken from Refs.\cite{Ruan:2018tsw,Luo:2025ewp}, respectively. The quantum-limit curve for future high-frequency gravitational-wave detectors is taken from Ref.~\cite{Guo:2025cza}.}
    \label{fig:gw}
\end{figure}

Before thermalization, as shown in~\cite{An:2026hiq} (see also in SUPP), in the IR region, $\tilde T_{ij}^{TT}(\eta,k)$ decays as $a^{-2}\sim\eta^{1+2\nu}$. Therefore, for modes with a conformal thermalization time $\eta_{\rm thermal} > k^{-1}$, the $\eta'$ integral in Eq.~\eqref{eq:hTT} yields the following IR behavior for the metric perturbation:
\bea
\tilde{h}_{ij}^{TT} \sim 
\begin{cases}
k^{-1}\ln k\,, & n=4 \\
k^{\frac{2}{n-2}-2}\,, & n>4 
\end{cases}\ .
\eea
Consequently, the IR part of the GW spectrum scales as
\bea\label{eq:omega IR scaling}
\Omega_{\rm GW}^{\rm IR}(k) \sim 
\begin{cases}
k^3 \ln^2 k\,, & n=4 \\
k^{\frac{n+2}{n-2}}\,, & n>4 
\end{cases}\ .
\eea
For modes with \(k\eta_{\rm thermal}<1\), \(\eta_{\rm thermal}\) acts as an effective cutoff for the \(\eta'\) integral in Eq.~\eqref{eq:hTT}. As a result, the \(\eta'\) integration does not introduce additional powers of \(k\), leading to
\[
\Omega^{\rm IR}_{\rm GW}(k) \sim k^{4-2\nu}.
\]
As illustrated in Fig.~\ref{fig:gw0}, for \(\alpha=0.05\), which saturates the PBH bound, the SM sector becomes dominant within about one decade of expansion after bubble collisions. The Universe then enters the radiation-dominated regime, and in the deep infrared one obtains \(\Omega_{\rm GW}\sim k^3\log^2 k\). By contrast, for \(\alpha=5\times10^{-4}\), the Universe remains dominated by \(\rho_{\rm DS}\) for more than \(10^3\) in scale-factor growth. Consequently, in the region \(k/H_c\sim 10^{-2}\), the IR slopes behave as \(k^2\) and \(k^{7/3}\) for \(n_{\rm DS}=5\) and \(6\), respectively.

Numerical simulations show that, in the transition region \(k/H_c\sim 0.1{-}1\), the spectrum scales approximately as \(\Omega_{\rm GW}\sim k^{1.5}\), while in the ultraviolet region it scales as \(\Omega_{\rm GW}\sim k^{-2}\). These features arise from the large Lorentz boost of the bubble walls and from the expansion history of the Universe. In practice, once the GW spectrum is measured, it can be compared with templates generated from numerical simulations to determine the details of the cosmic expansion history and the phase-transition parameters. A more detailed analysis of the GW spectrum is presented in the Supplemental Material.

To obtain the present-day GW spectrum, we include the redshift from the epoch of bubble collisions to today. This redshift is determined by the latent energy released into the SM sector and by the effective number of relativistic degrees of freedom after the SM sector thermalizes. In Fig.~\ref{fig:gw}, we show the resulting present-day GW spectra for several choices of \(\rho_c\) and \(\alpha\), assuming \(g_\star^{\rm SM}=100\).

\bigskip

\noindent{\it Summary and discussion} --- In this work, as illustrated in Fig.~\ref{fig:strategy}, we propose that the Big Bang expansion of our Universe may have originated from a vacuum decay triggered by the evolution of a DS. We show that the GWs produced during this phase transition exhibit distinctive features, which can help identify the nature of the transition and probe the details of the cosmic expansion history.

Collisions of highly boosted bubbles can produce heavy particles, such as right-handed neutrinos, whose decays may generate the baryon asymmetry through baryogenesis processes~\cite{Cataldi:2024pgt,Mansour:2023fwj}. In this scenario, baryogenesis may also occur during the thermalization of the high-energy shells. We leave a detailed investigation of this possibility to future work.

A similar scenario, in which the evolution of the Hubble rate triggers a vacuum decay within a DS, has been proposed in Refs.~\cite{Freese:2023fcr,Casey:2024jep}, with the associated GW properties systematically studied in Ref.~\cite{An:2026hiq}, and the late time cosmological consequences investigated in Ref.~\cite{Bai:2026sux}.



\bigskip

\noindent{\it Acknowledgement}--- We thank Zhen Liu for pointing out the collision processes that are not suppressed by large boost factor. We thank Haixing Miao for the information of high frequency GW proposals. This work is supported in part by the National Key R\&D Program of China under Grants Nos. 2021YFC2203100 and 2017YFA0402204, the NSFC under Grant Nos. 12475107 and 12525506.


\bigskip

\onecolumngrid
\newpage


\setcounter{equation}{0}
\renewcommand{\theequation}{S\arabic{equation}}

\setcounter{figure}{0}
\renewcommand{\thefigure}{S\arabic{figure}}

\begin{center}
  {\large{\textbf{\textit{Supplemental Material}}}}  
\end{center}

\section{Background evolution}

The energy density of the Universe during Standard Model (SM) phase transitions
induced by the dark sector (DS) can be decomposed into four components: the
dark-sector energy density, $\rho_{\rm DS}$; the average vacuum energy density,
$\rho_V$; the average energy density stored in the bubble walls before
collisions, $\rho_w$; and the average energy density carried by the
high-energy shells after bubble collisions, $\rho_s$. During the phase
transition, the bubble walls are accelerated by the vacuum-energy difference.
Energy conservation then gives
\begin{equation}
\label{eq:rhows}
\dot\rho_w+\dot\rho_s+4H\left(\rho_w+\rho_s\right)
= -\dot\rho_V \, .
\end{equation}
We assume that the dark-sector energy density redshifts as a power law,
\begin{equation}
\rho_{\rm DS} \propto a^{-n_{\DS}} \, .
\end{equation}
A representative benchmark is the dark-kination scenario, for which $n_\DS=6$.

The Hubble expansion rate is determined by
\begin{equation}
\label{eq:H2}
H^2 =
\frac{1}{3\mpl^2}
\left(
\rho_w+\rho_s+\rho_V+\rho_{\rm DS}
\right) .
\end{equation}

To estimate the false-vacuum fraction, we note that the bubble-wall velocity
rapidly approaches the speed of light after nucleation. The fraction of space
remaining in the false vacuum is therefore equal to the probability that no
bubble has nucleated within the past light cone of a given point. It can be
written as
\begin{equation}
\label{eq:fFV}
f_{\rm FV}(t)
=
\exp\left[
-\frac{4\pi}{3}
\int_{t_c}^{t} dt'\,
\Gamma(t')\,a^3(t')\,r^3(t,t')
\right] .
\end{equation}
Here $t$ denotes physical time, and $r(t,t')$ is the comoving radius at time
$t$ of a bubble nucleated at time $t'$. In terms of conformal time $\eta$, this
radius is
\begin{equation}
r(t,t')=\eta(t)-\eta(t') \, .
\end{equation}
Thus, the average vacuum-energy density during the phase transition is
\begin{equation}
\rho_V(t)=L f_{\rm FV}(t) \, ,
\end{equation}
where $L$ denotes the latent heat.

Using Eq.~\eqref{eq:rhows}, the total energy density in the bubble walls and
high-energy shells can be expressed as
\begin{equation}
\rho_w(\eta)+\rho_s(\eta)
=
a^{-4}(\eta)\,L
\int_0^\eta
\D\eta'\,
a^4(\eta')\,
\frac{\D f_{\rm FV}(\eta')}{\D\eta'} \, .
\end{equation}
Since both the SM sector and the dark sector participate in the dynamics, there
is in general no simple closed-form expression for $\rho_w+\rho_s$,
$\rho_{\rm DS}$, or $\rho_V$ at arbitrary conformal time $\eta$. Nevertheless,
these quantities can be obtained numerically by solving
Eqs.~\eqref{eq:rhows}, \eqref{eq:H2}, and \eqref{eq:fFV}. Once the evolution of
the energy densities is known, the corresponding dependence of the scale factor
$a$ on conformal time $\eta$ can be determined.

\section{Simulation of gravitational waves}\label{sec: simulation of GW}

To calculate the GWs, we adopt the following metric:
\begin{align}
    \D s^2 = a^2 \D \eta^2 - a^2\left(\delta_{ij}+h_{ij}\right)\D x^i\D x^j \, ,
\end{align}
where $h_{ij}$ denotes the spatial metric perturbation, and the GWs are described by the transverse-traceless components $h_{ij}^{TT}$ of $h_{ij}$. We further define
\[
\mathrm{h}_{+,\times}=a\, e_{+,\times}^{ij} h_{ij}^{TT} \, ,
\]
where $e_{+,\times}^{ij}$ are the polarization tensors. Then $\mathrm{h}_{+,\times}$ satisfy the linearized Einstein equation,
\begin{align}\label{eq:GW basic formula}
    \tilde{\mathrm{h}}_{+,\times}(\eta, \mathbf{k})=16 \pi G \int_{-\infty}^{\infty} \D \eta^{\prime} \, G\left(\eta,\eta^{\prime},\mathbf{k}\right)\, a\left(\eta^{\prime}\right)\, \tilde{T}_{+,\times}\left(\eta^{\prime}, \mathbf{k}\right)\, ,
\end{align}
where $G(\eta,\eta',k)$ is the Fourier-space Green's function, and $\tilde T_{+,\times}$ is the projection of the energy-momentum tensor onto the polarization tensors $e_{+,\times}^{ij}$.

\subsection{The Green's function}

The Green's function defined in Eq.~\eqref{eq:GW basic formula} satisfies
\begin{equation}
G^{\prime \prime}\left(\eta,\eta^{\prime},\mathbf{k}\right) + \left(k^2-\frac{a^{\prime \prime}}{a}\right) G\left(\eta,\eta^{\prime},\mathbf{k}\right) = 0 \, ,
\end{equation}
with the initial conditions
\begin{equation}
G(\eta^{\prime},\eta^{\prime})=0 \, , \qquad G^{\prime}(\eta^{\prime},\eta^{\prime})=1 \, ,
\end{equation}
where the prime denotes differentiation with respect to conformal time $\eta$.

If $\rho_{\rm DS}$ dominates throughout the GW production period, $G(\eta,\eta',k)$ can be solved analytically in terms of Bessel functions,
\begin{align}\label{eq:bessel}
    G(\eta,\eta',k)=\frac{\pi}{2}\sqrt{\eta\eta'}\left[J_{\nu}(k\eta')Y_{\nu}(k\eta)-J_{\nu}(k\eta)Y_{\nu}(k\eta')\right] \, ,
\end{align}
where $\nu = \left|\frac{n-6}{2(n-2)}\right|$. However, in the generic case, where multiple components contribute as in Eq.~\eqref{eq:H2}, no closed-form expression for $G(\eta,\eta',k)$ is available. In general, it can be written as
\begin{align}\label{eq:general greens function}
G(\eta,\eta',\mathbf{k})=y_1(\eta,\mathbf{k})y_2(\eta',\mathbf{k})-y_2(\eta,\mathbf{k})y_1(\eta',\mathbf{k}) \, ,
\end{align}
where $y_1$ and $y_2$ are two independent solutions of
\begin{align}\label{eq:dy}
    y^{\prime \prime}\left(\eta,\mathbf{k}\right) +\left(k^2-\frac{a^{\prime \prime}}{a}\right) y\left(\eta,\mathbf{k}\right) =0 \, ,
\end{align}
with initial conditions $y_1(\eta_0,\bk)=0$, $y'_1(\eta_0,\bk)=1$, and $y_2(\eta_0,\bk)=1$, $y'_2(\eta_0,\bk)=0$.

As will be seen from the PBH bound, vacuum energy never dominates the total energy density of the Universe. Consequently, the scale factor approximately follows a power law, although the corresponding power may vary with time. Thus, $a''/a$ in Eq.~\eqref{eq:dy} can be estimated qualitatively as $1/\eta^2$. It then follows that, in the limits $k\eta \gg 1$ or $k\eta' \gg 1$, both $y_1$ and $y_2$ are oscillatory functions of $k\eta$ and $k\eta'$, respectively, with approximately constant amplitudes. This behavior can be seen explicitly from the asymptotic form of the Bessel functions in Eq.~\eqref{eq:bessel}. In our numerical simulation, we solve Eq.~\eqref{eq:dy} numerically. Since we compute $\Omega_{\rm GW}^*$ only during radiation domination, we have $y_1(\eta,\bk)\rightarrow k^{-1/2}\cos(k\eta+\varphi)$ and $y_2(\eta,\bk)\rightarrow k^{-1/2}\sin(k\eta+\varphi)$, with $\varphi$ some phase related to the background details. When calculating $\Omega_{\rm GW}^*$, we average the GW signal over a time interval much longer than the GW period, and therefore make the replacements $\langle\sin^2 k\eta\rangle \rightarrow 1/2$, $\langle\cos^2 k\eta\rangle \rightarrow 1/2$, and $\langle\sin k\eta \cos k\eta \rangle \rightarrow 0$.

\subsection{Simulation of $T_{ij}^{\rm TT}$ using the bulk-flow model}

In this subsection, we describe our procedure for simulating the evolution of
$T_{ij}^{\rm TT}$ within the bulk-flow model. Our approach closely follows the
treatment of Ref.~\cite{Konstandin:2017sat}. In the present case, however, the
typical bubble size is comparable to the Hubble scale, and therefore the
expansion of the Universe must be taken into account.

Since the typical boost factor of the bubble wall is very large, $\gamma \sim \mpl/m$, where $m$ denotes the characteristic energy scale of the phase-transition system, the potential term in the Lagrangian is suppressed relative to the kinetic term. Consequently, during collisions, the highly boosted bubble walls pass through each other and form highly boosted, energetic shells. As discussed extensively in Ref.~\cite{An:2026hiq}, the energy of these shells remains localized. Moreover, since all shells propagate at approximately the speed of light, the field configuration retains a bubble-like structure even after collisions.

We describe the phase transition by a scalar field $\phi$. In the thin-wall approximation, the spatial components of the energy-momentum tensor for a single
bubble centered at the origin are given by
\begin{align}
    T_{ij} = \hat{x}_i \hat{x}_j \left(\partial_r \phi\right)^2 .
\end{align}
Its Fourier transform is therefore
\begin{align}
    \widetilde{T}_{ij}(\bk)
    =
    \int
    \hat{x}_i \hat{x}_j
    e^{-i \bk\cdot\bx}
    \left(\partial_r \phi\right)^2
    r^2 \sin\theta \,
    \mathrm{d}r\,\mathrm{d}\theta\,\mathrm{d}\varphi .
\end{align}
Choosing the wave vector $\bk$ to lie along the $\hat{z}$ direction, we have
$\bk\cdot\bx = k r \cos\theta$. Defining $\zeta \equiv \cos\theta$ and applying
the thin-wall approximation, the above expression becomes
\begin{align}
    \widetilde{T}_{ij}(k)
    =
    \int
    \hat{x}_i \hat{x}_j
    e^{-i k r_s \zeta}
    \,\mathrm{d}\zeta\,\mathrm{d}\varphi
    \left[
    r_s^2
    \int_0^{r_s}
    \left(\partial_r \phi\right)^2
    \mathrm{d}r
    \right],
\end{align}
where $r_s$ denotes the bubble radius.

We then project $\widetilde{T}_{ij}$ onto the two tensor polarization components.
For $\bk \parallel \hat{z}$, these are given by
\begin{align}
    \widetilde{T}_{+}(k)
    &=\frac{1}{2}
    \int
    e^{-i k r_s \zeta}
    \left(1-\zeta^2\right)
    \cos 2\varphi\,
    \mathrm{d}\zeta\,\mathrm{d}\varphi
    \left[
    r_s^2
    \int_0^{r_s}
    \left(\partial_r \phi\right)^2
    \mathrm{d}r
    \right],\label{eq: tij bulk 1}\\
    \widetilde{T}_{\times}(k)
    &=
    \frac{1}{2}
    \int
    e^{-i k r_s \zeta}
    \left(1-\zeta^2\right)
    \sin 2\varphi\,
    \mathrm{d}\zeta\,\mathrm{d}\varphi
    \left[
    r_s^2
    \int_0^{r_s}
    \left(\partial_r \phi\right)^2
    \mathrm{d}r
    \right]\label{eq: tij bulk 2}\ .
\end{align}

We further define the radial integral 
\begin{align}
    I_r(\zeta, \varphi)=r_s^2\int_0^{r_s}\left(\partial_r\phi\right)^2 \mathrm{d}r \ .
\end{align}
$I_r(\zeta,\varphi)$ can be interpreted as the bubble wall's comoving energy per solid angle. For a bubble configuration (before or after the first collision), the total physical energy of the bubble can be written as $E_b = a \int I_r d\Omega$. $I_r$ can be simplified using the EoM of the scalar field.


Before collisions, the bubble walls are spherically symmetric, so that $E_b = a \int I_r d\Omega=4\pi a I_r$. The wall energy redshifts like radiation because the wall is ultra-relativistic, while it also absorbs energy from the false vacuum. The evolution equation for the total physical energy $E_b$ can be written as
\begin{align}
    \frac{\D E_b}{\D t}+HE_b=4\pi a^2 r_s^2 L\ .
\end{align}
We then express $E_b$ in terms of $I_r$ and rewrite the evolution equation in conformal time. The solution then gives
\begin{align}\label{eq:Ireta}
    I_r(\eta)=\frac{L}{a^2(\eta)}\int^{\eta}_{\eta_0} a^4(\eta')r_s^2(\eta')\D \eta'\ ,
\end{align}
where $\eta_0$ denotes the conformal time when the bubble is nucleated. 

After bubble collisions, the spherical symmetry of the wall’s comoving energy per solid angle is broken. The energy shells after collision evolve according to the bulk flow model discussed previously. The bubble wall then stops gaining energy at the collision time $\eta_{coll}$ and subsequently decays like radiation, which implies that
\begin{align}\label{eq: I_r}
    I_r(\eta,\zeta,\varphi)=\frac{a^2(\eta_{coll})}{a^2(\eta)} I_r(\eta_{coll},\zeta,\varphi)=\frac{L}{a^2(\eta)}\int^{\eta_{coll}(\zeta,\varphi)} _{\eta_0} a^4(\eta')r_s^2(\eta')\D \eta'\ .
\end{align}
Here we introduce a dependence of $\eta_{coll}$ on $\zeta$ and $\varphi$, since the collision time varies with direction. 
Combined with Eq.~\eqref{eq: tij bulk flow}, we obtain the contribution of a single bubble to $\tilde{T}_{ij}$. Summing over all bubbles then gives the total Fourier modes of the energy-momentum tensor:
\begin{align}\label{eq: total tij}
    \tilde{T}^{\rm tot}_{+,\times}(\mathbf{k})=\sum_n e^{ik_z z_n} \tilde{T}_{+,\times}^n(\mathbf{k})\ ,
\end{align}
where $z_n$ is the z coordinate of the center of the $n$th bubble. 


\section{GW shape analysis}
In this section, we give a brief semi-analytical description of the shape of the GW spectrum from a dark-sector-induced phase transition and compare it with simulation results.
The GW spectral density is determined by the Green's function $y(\eta, \mathbf{k})$ and the source term $\tilde{T}(\eta, \mathbf{k})$. Here, $y(\eta, \mathbf{k})$ encodes the equation of state of the universe, while $\tilde{T}(\eta, \mathbf{k})$ is governed by the evolution of bubble shells. We therefore expect that the transition points in the background and bubble evolution determine the positions of the turning points in the spectral density shape.
Before we proceed to the calculation of the spectral density shape, let us first discuss several critical times in the background and bubble evolution.
\begin{itemize}

    \item $\eta_c$: the conformal time of the phase transition. We define it as the conformal time when $\Gamma/\mathcal{V}=H^4$ is satisfied. Since the phase transition typically lasts for about one e-fold, this time also sets the characteristic timescale of the transition.

    \item $\eta_{eq}$: the dark-sector-SM equality time which satisfies $\rho_{\DS}(\eta_{eq})=\rho_{\mathrm{SM}}(\eta_{eq})$. This time marks the turning point in the equation of state of the cosmological background.
    
    \item $\eta_{T}$: the thermalization time. This time marks the termination of the GW source.
\end{itemize}

Since thermalization takes roughly $\mathcal{O}(10^2)\sim\mathcal{O}(10^3)$ e-folds, we expect $\eta_c\ll \eta_{T}$. In our parameter space, the SM sector is initially subdominant when the phase transition starts, indicating that $\eta_{eq}\gtrsim \eta_c$. If $\eta_{eq}>\eta_T$, the latent heat is small and suppresses the GW signal. We therefore focus mainly on the parameter region where $\eta_c\lesssim\eta_{eq}<\eta_{T}$.

\subsection{theoretic predictions}

The energy density of GW can be expressed as
\begin{align}\label{eq:gw density}
    \rho_{GW}&=\frac{1}{16\pi G a^4 V}\sum_{A=+,\times}\int\frac{\D^3 k}{(2\pi)^3}|\tilde{\mathrm{h}}'_{A}(k)|^2\ .
\end{align}
Combined  with Eq.~\eqref{eq:GW basic formula} and Eq.~\eqref{eq:general greens function}, Eq.~\eqref{eq:gw density} can then be expressed as
\begin{align}
    \rho_{GW}=\frac{1}{a^4 \mpl^2 V} \int \frac{k^3 \mathrm{d} k}{2 \pi^2} &\sum_{A=+, \times}\left[\left|\int_{-\infty}^{\infty} \mathrm{d} \eta^{\prime} y_1\left(\eta^{\prime}, \mathbf{k}\right) a\left(\eta^{\prime}\right) \tilde{T}^{\rm tot}_{A}\left(\eta^{\prime}, \mathbf{k}\right)\right|^2 \right.\nonumber\\
    &\left.+\left|\int_{-\infty}^{\infty} \mathrm{d} \eta^{\prime} y_2\left(\eta^{\prime}, \mathbf{k}\right) a\left(\eta^{\prime}\right) \tilde{T}^{\rm tot}_{A}\left(\eta^{\prime}, \mathbf{k}\right)\right|^2\right]\ .
\end{align}
Here we use the relation that $\langle y_1'^2(\eta,\mathbf{k})\rangle\rightarrow k/2$, $\langle y_2'^2(\eta,\mathbf{k})\rangle\rightarrow k/2$ and $\langle y_1'(\eta,\mathbf{k})y_2'(\eta,\mathbf{k})\rangle\rightarrow 0$ when $k\eta\rightarrow\infty$. We further define the GW spectral density, which reads
\begin{align}\label{eq:GW spectral density}
    \Omega_{GW}=\frac{\D \rho_{GW}}{\rho_{SM}\D\ln k}&= C_{GW} k^4\sum_{A=+, \times}\left[\left|\int_{-\infty}^{\infty} \mathrm{d} \eta^{\prime} y_1\left(\eta^{\prime}, \mathbf{k}\right) a\left(\eta^{\prime}\right) \tilde{T}^{\rm tot}_{A}\left(\eta^{\prime}, \mathbf{k}\right)\right|^2 \right.\nonumber\\
    &\left.+\left|\int_{-\infty}^{\infty} \mathrm{d} \eta^{\prime} y_2\left(\eta^{\prime}, \mathbf{k}\right) a\left(\eta^{\prime}\right) \tilde{T}^{\rm tot}_{A}\left(\eta^{\prime}, \mathbf{k}\right)\right|^2\right]\ ,
\end{align}
with $C_{GW} = [2\pi^2 a^4 \mpl^2 V(\rho_w+\rho_s)]^{-1}$. 
Here the denominator of $C_{GW}$ is $\rho_w+\rho_s$ instead of the total energy of the universe since we want to compare the GW density with SM radiation. 

Next, we discuss the shape of $\Omega_{GW}$. In general, $\Omega_{GW}$ can be divided into four types of regions:
\begin{itemize}
    \item $\mathrm{I}$: Deep IR region, corresponding to $k\ll \eta_{T}^{-1}$.
    In region $\mathrm{I}$, the GW source has vanished after thermalization. The spectrum therefore follows the $k^3$ scaling imposed by causality in this region~\cite{Caprini:2009fx,Cai:2019cdl}. If $\eta_T<\eta_{eq}$, the spectral shape is expected to approach $k^{4-\nu}$ in the deep IR for $k\ll\eta_{T}^{-1}$~\cite{Domenech:2020kqm}. However, the GW signal is extremely weak there, and its IR part is difficult to detect; we therefore do not consider this situation.
    
    \item $\mathrm{II}$: RD free-streaming region, corresponding to $\mathcal{O}(\eta_{T}^{-1})< k< \mathcal{O}(\eta_{eq}^{-1})$. This region appears only when $\eta_{eq}<\eta_{T}$.

    \item $\mathrm{III}$: Dark-sector free-streaming region, corresponding to $\mathcal{O}(\eta_{eq}^{-1})< k< \mathcal{O}(\eta_{c}^{-1})$. This region appears only when $\eta_{eq}>\eta_{c}$.

    \item $\mathrm{IV}$: Flat-spacetime region, corresponding to $k> \mathcal{O}(\eta_{c}^{-1})$.
    During the phase transition, the GW source peaks when bubble collisions become efficient, roughly around $\eta_c$. For modes with $k\eta_c \gg 1$, the Hubble expansion is negligible. We therefore expect that for $k\gg\eta_c^{-1}$, the shape of $\Omega_{GW}$ is similar to that in flat spacetime. The bulk flow model in flat spacetime predicts a $k^1$ scaling in the IR and a $k^{-2}$ scaling in the UV for $\Omega_{GW}$~\cite{Konstandin:2017sat,Ellis:2020nnr,Jinno:2017fby}.
\end{itemize}

Therefore, the only regions that remain unknown are $\mathrm{II}$ and $\mathrm{III}$, which correspond to the bulk flow model in the presence of a cosmological background. Both regions satisfy $k<\mathcal{O}(\eta_c^{-1})$. For convenience, we will adopt $k\ll\eta_c^{-1}$ in the following discussion. In this limit, the phases in Eq.\eqref{eq: total tij} approach unity. Thus, we can use the single-bubble contribution to $\tilde{T}_{ij}$ when discussing the $k$-dependence of $\Omega_{GW}(k)$.
Using Eq.~\eqref{eq:GW spectral density}, the spectral shape of $\Omega_{GW}$ is determined by the following integral:
\begin{align}
    I_{a}^A(k)=\int_{-\infty}^{\infty} \mathrm{d} \eta^{\prime} y_a\left(\eta^{\prime}, \mathbf{k}\right) a\left(\eta^{\prime}\right) \tilde{T}_A\left(\eta^{\prime}, \mathbf{k}\right)\ , 
\end{align}
where $a=1,2$ and $\tilde{T}_A$ is given by Eq.\eqref{eq: tij bulk 1} and Eq.\eqref{eq: tij bulk 2}. 
Suppose that the background evolution follows $\rho\propto a^{-n_\DS}$, the mode function $y_a\left(\eta, \mathbf{k}\right)$ yields
\begin{align}
    y_1\left(\eta, \mathbf{k}\right)=\sqrt{\frac{\pi \eta}{2}}J_{\nu}(k\eta),\qquad y_2\left(\eta, \mathbf{k}\right)=\sqrt{\frac{\pi \eta}{2}}Y_{\nu}(k\eta)\ ,
\end{align}
with $\nu=\left|\frac{n_\DS-6}{2(n_\DS-2)}\right|$. 
We define $\eta_f$ as the conformal time when bubble collisions finish. Typically, $\eta_f/\eta_c\sim\mathcal{O}(1)$.
We separate $I_{a}^A(k)$ into two parts:
\begin{align}
    I_{a}^A(k)&=I_{a,\eta_f}^A(k) + I_{a,\infty}^A(k)\nonumber\\
    &=\int_{-\infty}^{\eta_f} \mathrm{d} \eta^{\prime} y_a\left(\eta^{\prime}, \mathbf{k}\right) a\left(\eta^{\prime}\right) \tilde{T}_A\left(\eta^{\prime}, \mathbf{k}\right) + \int_{\eta_f}^{\infty} \mathrm{d} \eta^{\prime} y_a\left(\eta^{\prime}, \mathbf{k}\right) a\left(\eta^{\prime}\right) \tilde{T}_A\left(\eta^{\prime}, \mathbf{k}\right)\ , 
\end{align}
Here, $I_{a,\eta_f}^A(k)$ represents the contribution before and after $\eta_f$. In evaluating this term, since $k\ll\eta_c^{-1}$ we can neglect the $k$ dependence in $\tilde{T}_A\left(\eta, \mathbf{k}\right)$ and expand $y_a(\eta,\mathbf{k})$ for small $k$, yielding
\begin{align}
    I_{a,\eta_f}^A(k)&\propto\int_{-\infty}^{\eta_f} \mathrm{d} \eta^{\prime} \sqrt{\eta'}\left(k\eta'\right)^{\pm \nu}a\left(\eta^{\prime}\right) \tilde{T}_A\left(\eta^{\prime}, \mathbf{0}\right) \propto k^{\pm\nu}\ .
\end{align}
$I_{1,\eta_f}^A(k)\propto k^{\nu}$ can be neglected compared to $I_{2,\eta_f}^A(k)\propto k^{-\nu}$. Therefore, we only need to consider $I_{2,\eta_f}^A(k)$.

Next, we consider $I_{a,\infty}^A(k)$. Since the phase transition ends at $\eta_f$,  $\tilde{T}_A(\eta,k)$ can be expressed as
\begin{align}
    \tilde{T}_A(\eta,k) &= \frac{1}{2} a^{-2}(\eta) \int e^{-i k \eta \zeta}\left(1-\zeta^2\right) \left\{\begin{array}{c}
         \cos 2\varphi  \\
          \sin 2\varphi
    \end{array}\right\} a^2(\eta_c)I_r(\eta_c, \zeta, \varphi)\mathrm{d} \zeta \mathrm{~d} \varphi \ .
\end{align}
Here, we make use of the relation in Eq.\eqref{eq: I_r}. 
For convenience, we define the $\zeta$-dependent part as $f(\zeta)$.
Therefore, $I_{a,\infty}^A(k)$ reduces to
\begin{align}\label{eq: bulk flow IR}
    I_{a,\infty}^A(k)&\propto\int_{\eta_f}^{\infty} \frac{\mathrm{d} \eta^{\prime}}{\eta'^{\nu}} \left\{\begin{array}{c}
         J_{\nu}(k\eta)  \\
          Y_{\nu}(k\eta)
    \end{array}\right\} \int e^{-i k \eta \zeta}f(\zeta)\mathrm{d} \zeta\nonumber\ ,\\
    &\propto \begin{cases}
k^{-\frac{1}{2}}\ln k\eta_f\,, & \nu=\frac{1}{2} \\
k^{\nu-1}\,, & 0<\nu <\frac{1}{2}
\end{cases}\ .
\end{align}

Here we retain only the leading contribution in the limit $k\eta_f\ll 1$. This result can be understood as follows. The oscillation of the Green's function for $k\eta\gg 1$, combined with $\tilde{T}_A$, introduces an effective cutoff $\eta_{cut}\sim k^{-1}$ in the integral in Eq.\eqref{eq: bulk flow IR}. Consequently, $I_{a,\infty}^A(k)\sim k^{\nu-1} \int_{k\eta_f}^{\mathcal{O}(1)} \mathrm{d} x/x^{\nu+1/2}$, which, by counting powers of $k$, recovers the results in Eq.\eqref{eq: bulk flow IR}. This indicates that the integral in Eq.\eqref{eq: bulk flow IR} is dominated by the contribution from large $\eta$. Therefore, if the equation of state of the universe changes at $\eta_{eq}$, we expect that for $k>\eta_{eq}^{-1}$ the exponent $\nu$ takes the value corresponding to dark-sector domination, while for $k<\eta_{eq}^{-1}$ it takes the RD value.  
This predicts a turning point in the GW spectral shape around $k\sim\eta_{eq}^{-1}$.

Compared with $I_{a,\eta_f}^A(k)$, $I_{a,\infty}^A(k)$ is always flatter in the regime $k\eta_f\ll 1$. In the bulk flow model, $I_{a,\infty}^A(k)$ therefore dominates the IR scaling. 
For the GW spectral density, we thus obtain
\begin{align}
    \Omega_{GW}\propto k^4 |I_{a,\infty}^A(k)|^2\propto \begin{cases}
k^{3}\ln^2 k\eta_f\,, & \nu=\frac{1}{2} \\
k^{2+2\nu}\, & 0<\nu <\frac{1}{2}
\end{cases}\ .
\end{align}

\subsection{simulation results}
\begin{figure}[htb]
    \centering
    \includegraphics[width=\linewidth]{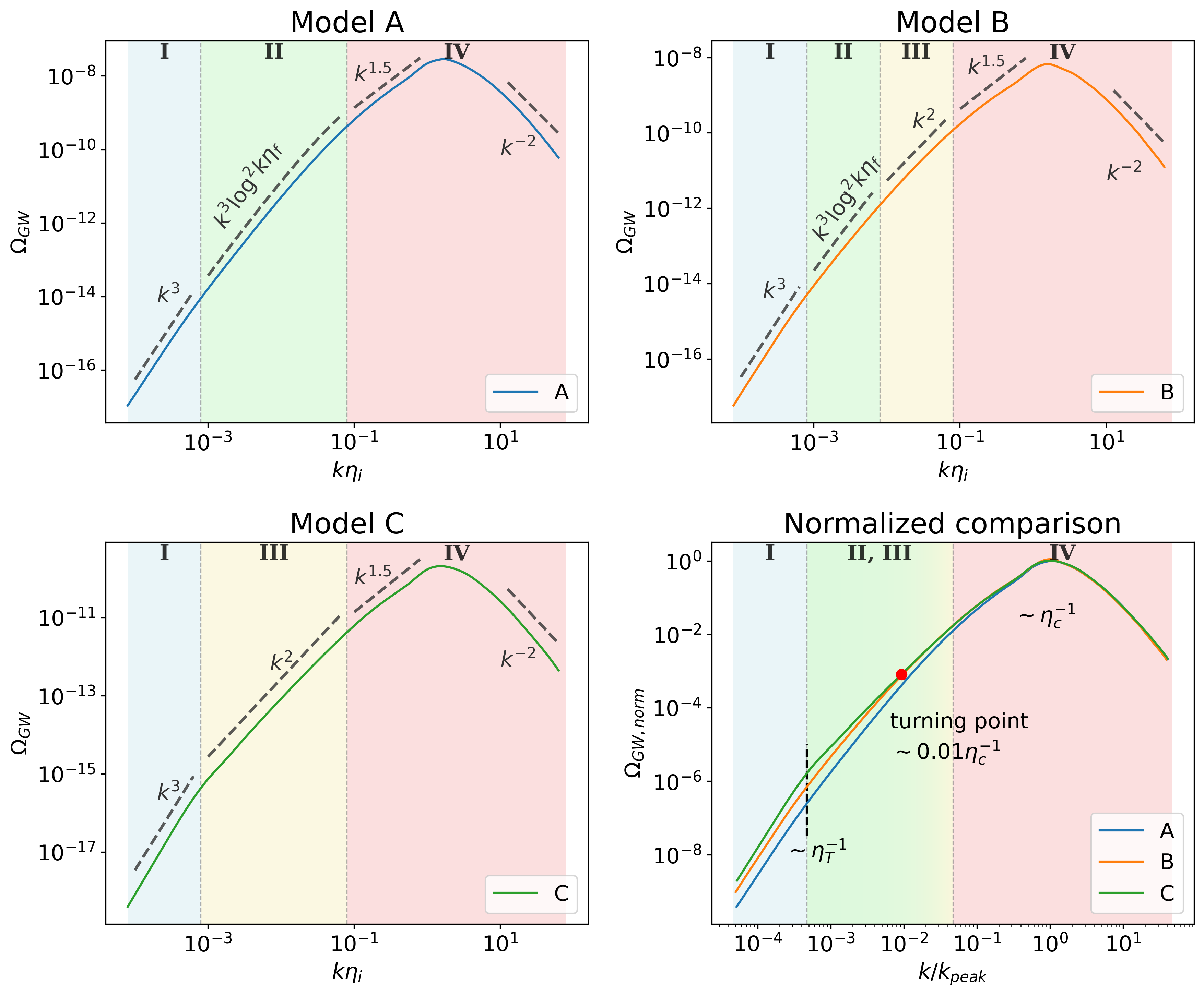}
    \caption{Models A, B, and C correspond to the simulation results with $\eta_{eq}/\eta_c=2.75, 12, 375$, respectively. In model A, the IR spectrum exhibits a $k^3\log^2 k\eta_f$ scaling. In model C, the IR spectrum follows a $k^2$ scaling. In model B, the IR scaling displays a turning point around $k\sim 0.01\eta_c^{-1}$. A comparison of the three models is shown in the bottom-right panel. We use different colors to represent four types of regions. }
    \label{fig:omega gw}
\end{figure}

In our numerical simulation, we adopt $n_D=6$. We choose three sets of parameters with the same $\eta_c$ and $\eta_T$ but different $\eta_{eq}$ by adjusting the value of $\alpha = L/\rho_{tot}$. We model the effect of thermalization as an exponential decay of the source on a timescale $\eta_T$, given by
\begin{align}
    \tilde{T}_{ij}^{Thermal}=\tilde{T}_{ij}e^{-\eta/\eta_T}
\end{align}
We set $\eta_T/\eta_c = 2.5\times 10^3$ in all three models. 
For each parameter set, we generate 160 realizations of the nucleation history, each containing about 100 bubbles, and compute the average $\Omega_{GW}$ following Sec.~\ref{sec: simulation of GW}.
The results are presented in Fig.~\ref{fig:omega gw}. 
For comparison, we also show the normalized GW spectra of the three models in the bottom-right panel.

For model A, we choose $\eta_{eq}/\eta_c=2.75$. In this case, $\eta_{eq}\approx \eta_{f}$, so the universe enters the RD era around the time the phase transition ends. We therefore expect the IR part of the GW spectral density to scale as $k^3\log^2k\eta_f$ down to $k\sim \eta_T^{-1}$, as indicated by the green shaded region in the upper-left panel of Fig.~\ref{fig:omega gw}.

For model B, shown in the upper-right panel of Fig.~\ref{fig:omega gw}, we set $\eta_{eq}/\eta_c=12$. Here a turning point in the IR scaling is expected. In the simulation result, we find this turning point to be located at $k\sim 0.01\eta_c^{-1}$. For $k>0.01\eta_c^{-1}$, $\Omega_{GW}\propto k^2$, while for $\eta_T^{-1}<k<0.01\eta_c^{-1}$, $\Omega_{GW}\propto k^3\log^2k\eta_f$. 
This turning point is more clearly visible in the comparison plot.

For model C, we further lower $\alpha$ and choose $\eta_{eq}/\eta_c=375$. Since now $\eta_{eq}$ is closer to the thermalization time $\eta_T$, we expect the dark-sector free-streaming region to become wider. As shown in the bottom-left panel of Fig.~\ref{fig:omega gw}, almost the entire IR part now belongs to region $\mathrm{III}$, which scales as $k^2$.

For the spectra in the red shaded regions, the influence of the background evolution is expected to be negligible. In the deep UV, $\Omega_{GW}\propto k^{-2}$, which matches the scaling of the flat-spacetime bulk flow model. In the IR, $\Omega_{GW}\propto k^{1.5}$, which is steeper than the $k^1$ scaling predicted by the flat-spacetime bulk flow model. This deviation may result from corrections induced by the Hubble friction. The spectrum in the blue shaded region corresponds to the causal regime, scaling as $k^3$ as expected.

\section{PBH production during SM sector vacuum decay induced by the evolution of a power law dark sector}

During a first-order phase transition, the stochastic nature of bubble nucleation implies that some patches of false vacuum decay later than others. When the regions surrounding a false-vacuum domain (FVD) transition to a vacuum with lower energy density, the energy density difference between the surrounding region and the FVD redshifts as \(a^{-4}\), where \(a\) is the scale factor of the FRW Universe. As a result, an overdensity develops between the FVD and its surroundings. Once this density contrast becomes sufficiently large, the FVD begins to collapse and can eventually form a black hole (BH), usually referred to as a primordial black hole (PBH). 

\begin{figure}[htp]
    \centering
    \begin{minipage}{0.48\textwidth}
        \centering
        \includegraphics[width=\textwidth]{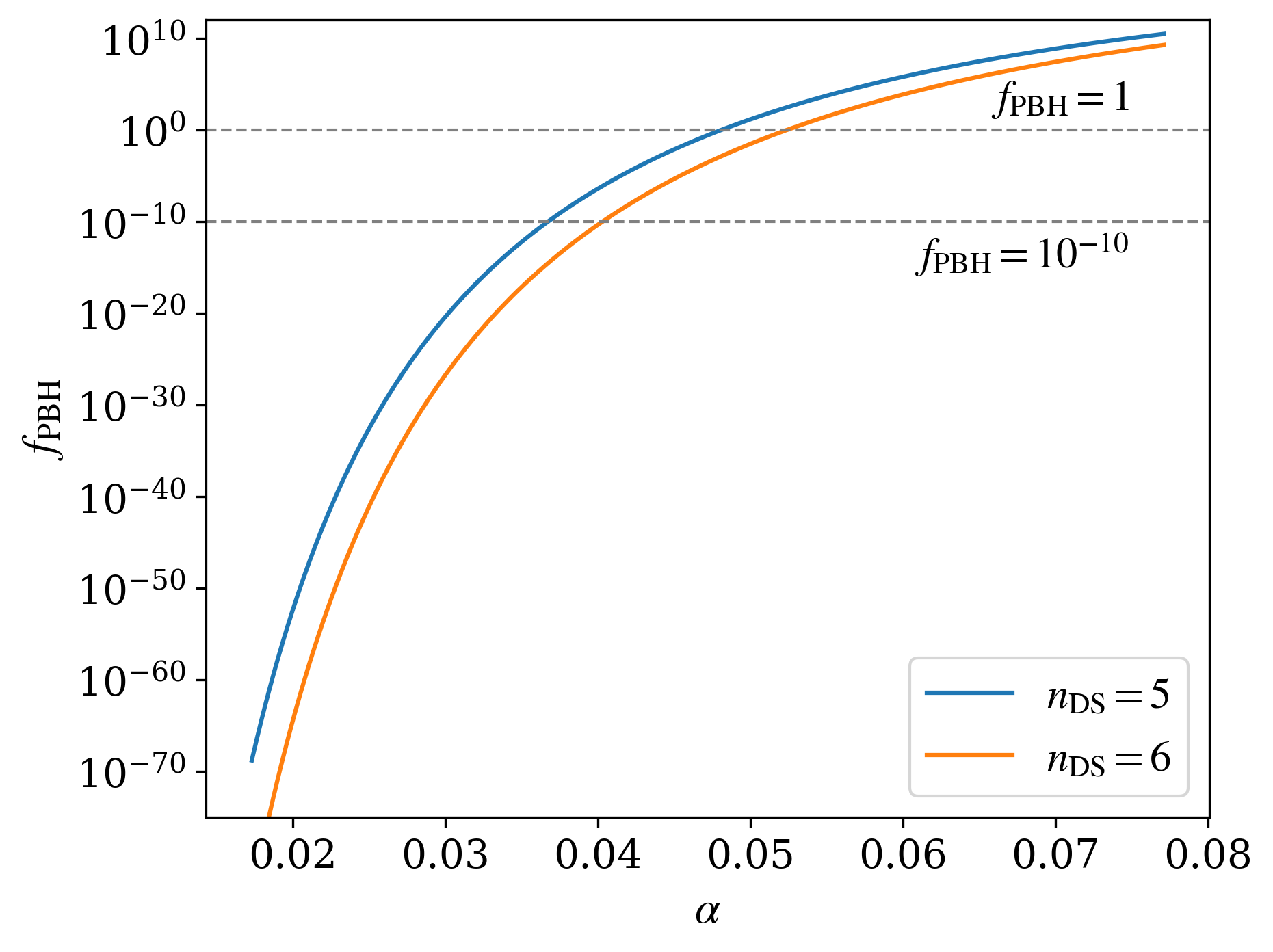}
        \label{fig:sub1}
    \end{minipage}
    \hfill
    \begin{minipage}{0.48\textwidth}
        \centering
        \includegraphics[width=\textwidth]{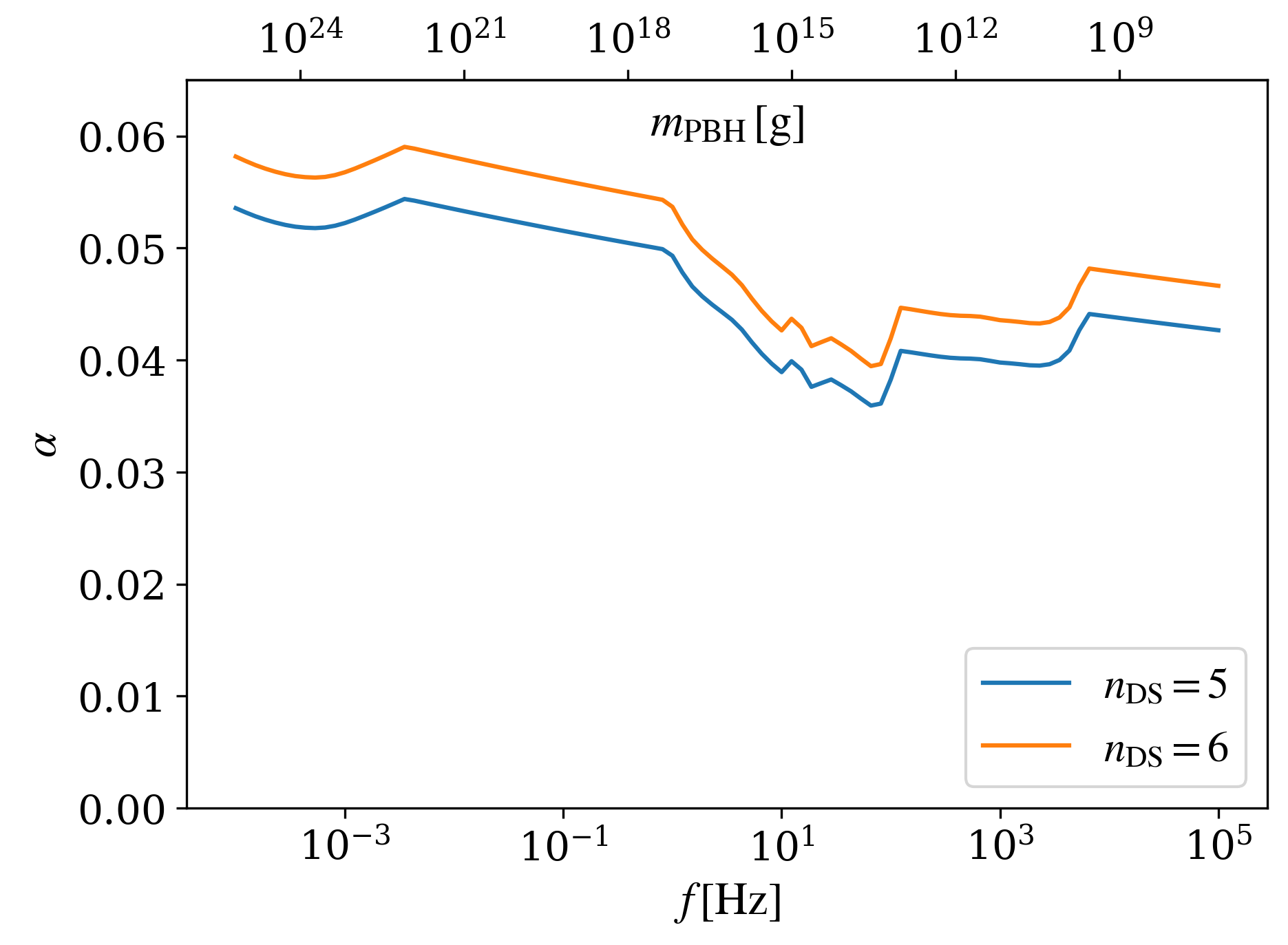}
        \label{fig:sub2}
    \end{minipage}
    \caption{The left panel shows the dependence of $f_{\mathrm{PBH}}$ on $\alpha$. As a benchmark, we consider a phase transition that produces GWs peaked at $f=10\mathrm{Hz}$. The right panel displays the constraints on $\alpha$ for different PBH masses. 
    We adopt the constraints on $f_{\mathrm{PBH}}$ from~\cite{Carr:2020gox,Green:2020jor,Keith:2020jww,Acharya:2020jbv,Clark:2018ghm,Carr:2009jm,Laha:2020ivk,Croon:2020ouk,EROS-2:2006ryy,Niikura:2019kqi}. 
    We also indicate the corresponding peak frequency of the GWs from the phase transition. }
    \label{fig:both}
\end{figure}

Extensive searches for PBHs have been conducted across a wide range of masses. In this appendix, we estimate the number density of PBHs in the metastable SM vacuum decay scenario and derive the corresponding constraints on its parameter space.
There are mainly two criteria for determining PBH formation. The local density contrast criterion suggests that a PBH forms when the density contrast exceeds a critical value $\delta_c$~\cite{Liu:2021svg,Hashino:2021qoq,Kawana:2022olo}. The Schwarzschild criterion, on the other hand, requires the Schwarzschild radius of the region to exceed its size~\cite{Flores:2024lng,Baker:2021sno,Jung:2021mku,Ning:2026nfs}. 
In this paper, we use the Schwarzschild criterion method from Ref.~\cite{Flores:2024lng,Lewicki:2023ioy} to estimate the PBH density. 
As we will see, the constraint on $\alpha$ is insensitive to the PBH abundance; therefore, the choice of criterion does not significantly affect the constraints. 
In this method, the shapes of the FVDs with the largest contrasts are approximated as nearly spherical, and the contrast mass is given by
\bea
\Delta m = \frac{4}{3}\pi \Delta V a^3 r^3 + 4\pi a^2 r^2 \sigma \bar\gamma \ ,
\eea 
where $r$ is the comoving radius of the FVD, $\Delta V = L - (L f_{\rm FV}+\rho_{R,SM})$ with $L$ the latent energy density released during the phase transition and $\rho_{R,sm}$ the average radiation energy density in the SM sector. $\sigma$ is the tension of the bubble wall in its rest frame and $\bar\gamma$ is the average boost factor of the bubble walls. Thus, the product $\sigma\bar\gamma$ is the tension of the average tension surrounding the wall, which can be estimated as 
\bea
\sigma \bar\gamma(\eta) = \frac{\sum E_b}{\sum A_{b}}=\frac{a^{-1} \int d\eta' n(\eta') I_r(\eta)}{\int d\eta' n(\eta') (\eta-\eta')^2} \ ,
\eea
where $A_{bub}$ is the physical area of bubble walls, $I_r(\eta)$ is given in Eq.~\eqref{eq:Ireta} and $n(\eta') = a^4(\eta') \frac{\Gamma}{\cal V} P(\eta')$. Then, the Schwarzschild criterion $r_{\rm phys} < 2 G \Delta m$ for PBH to form  requires~\cite{Lewicki:2023ioy}
\begin{align}\label{eq:schwarszchild}
    ar<2G \left(\frac{4}{3}\pi\Delta Va^3r^3+4\pi a^2 r^2\sigma \bar{\gamma}\right) \ ,
\end{align}
which consequently requires 
\begin{align}
    a r \geq \sqrt{\left(\frac{\sigma \bar \gamma}{2\Delta V}\right)^2+\frac{\mpl^2}{3\Delta V}} - \left(\frac{\sigma \bar \gamma}{2\Delta V}\right) \ . 
\end{align}
We define this quantity as $ar_{low}$. 
The lower bound for PBH formation decreases with time because the average tension of the bubble wall increases during the phase transition. 
Furthermore, for a PBH to form, one must also account for the expansion of the Universe, requiring the comoving radius $r$ to be smaller than the comoving Hubble radius $R_H=(aH)^{-1}$~\cite{Garriga:2015fdk}. 
Consequently, at a given time $t$, FVDs with comoving radii satisfying $r_{low}<r<1/(aH)$ will collapse into PBHs. 
At time $t$, the newly formed PBHs correspond to those regions whose radii cross either the Hubble radius or the lower bound $r_{low}$. 
Hence, the PBH energy density can then be expressed as
\begin{align}
    \rho_{\mathrm{PBH}}(t)=a^{-3}(t)\int_{t_\times}^t\D t' a^3(t')\left[\frac{\D R_H}{\D t'}m(R_H, t')\frac{\D n_{\mathrm{FVD}}(R_H, t')}{\D r} \right.\nonumber\\
    \left.-\frac{\D r_{low}}{\D t'}m(r_{low}, t')\frac{\D n_{\mathrm{FVD}}(r_{low}, t')}{\D r}\right]\ .
\end{align}
Here $t_{\times}$ refers to the time when $r_{low}=R_H$ and $\D n_{\mathrm{FVD}}(r,t)/\D r$ is the distribution function of FVDs at time $t$. $\D n_{\mathrm{FVD}}(r,t)/\D r$ counts the physical number density of FVDs per radius. In order to obtain $\D n_{\mathrm{FVD}}(r,t)/\D r$, we first define the function $f_{\text {FVD }}(r, t)$ which represents the probability of finding a spherical false vacuum with radius $r$ at time $t$.  
\begin{align}
    f_{\text {FVD }}(r, t)=\exp \left[-\frac{\Gamma}{\mathcal{V}} \int_0^t \mathrm{~d} t^{\prime} \frac{4 \pi}{3}\left(r+R\left(t, t^{\prime}\right)\right)^3 a^3\left(t^{\prime}\right)\right]\ .
\end{align}
Then the distribution function of FVDs can be expressed as ~\cite{Lewicki:2023ioy,Kodama:1982sf}
\begin{align}
    \frac{\D n_{\mathrm{FVD}}(r,t)}{\D r}\approx a^{-3}\frac{1}{8\pi}\frac{\partial^4}{\partial r^4} f_{\text {FVD }}(r, t)\ .
\end{align}
We use $f_{\mathrm{PBH}}=(\rho_{\mathrm{PBH}}/\rho_{m})_{\mathrm{today}}$ to evaluate the abundance of PBHs. The dependence of $f_{\mathrm{PBH}}$ on $\alpha=L/\rho_{\mathrm{tot}}$ is shown in Fig.\ref{fig:sub1}. The PBH abundance is very sensitive to the initial energy fraction $\alpha$. Based on the constraints on $f_{\mathrm{PBH}}$ in different mass ranges, we obtain the constraints on $\alpha$ for different PBH masses. We also translate the PBH mass into the corresponding peak frequency of the GWs from the phase transition. 
Our results imply that the upper bound on $\alpha$ is around $0.05$, and it depends only weakly on frequency.

\bibliographystyle{arxivref}
\bibliography{refs}



\end{document}